\documentclass[rmp,aps,floats,floatfix,superscriptaddress,nofootinbib,citeautoscript,twocolumn]{revtex4}
\usepackage{epsfig}
\usepackage{graphicx}
\usepackage{amssymb,amsmath}
\usepackage{float}
\usepackage{version}
\usepackage{color}
\usepackage{times}

\bibpunct{[}{]}{,}{n}{}{,}

\newcommand{\dd}{\textrm{d}}

\newcommand{\ee}{\text{e}}
\newcommand{\bv}{{\bf v}}

\newcommand{\bs}{\widehat{\boldsymbol \sigma}}

\newcommand{\cP}{{\cal P}}
\newcommand{\cN}{{\cal N}}

\newcommand{\hf}{{\widehat f}}
\newcommand{\Mvariable}{{}}
\newcommand{\Mfunction}{{}}
\newcommand{\p}{\partial}
\newcommand{\PFFT}{P_{_{\text{FFT}}}}
\graphicspath{{figures/}}

\begin{document}
\title{\textrm{Non Poissonian statistics in a low density fluid}} 

\date{\today}
\author{\textrm{\bf Paolo Visco}}

\affiliation{\textrm{Universit\'e Paris-Sud, LPTMS, UMR 8626, Orsay Cedex,
F-91405 and CNRS, Orsay, F-91405}}

\author{\textrm{\bf Fr\'ed\'eric van Wijland}}

\affiliation{\textrm{Laboratoire Mati\`ere et Syst\`emes Complexes (CNRS UMR
  7057), Universit\'e Denis Diderot (Paris VII), 10 rue Alice Domon et
  L\'eonie Duquet, 75205 Paris cedex 13, France}}

\author{\textrm{\bf Emmanuel Trizac}}

\affiliation{\textrm{Universit\'e Paris-Sud, LPTMS, UMR 8626, Orsay Cedex,
   F-91405 and CNRS, Orsay, F-91405}}

\begin{abstract}
  Our interest goes to the collisional statistics in an arbitrary
  interacting fluid.  We show that {\em even in the low density limit}
  and contrary to naive expectation, the number of collisions
  experienced by a tagged particle in a given time does not obey
  Poisson law, and that conversely, the free flight time distribution
  is not a simple exponential. As an illustration, the hard sphere
  fluid case is worked out in detail. For this model, we quantify
  analytically those deviations and successfully compare our
  predictions against molecular dynamics simulations.
\end{abstract}

\maketitle


The purpose of the present paper is to address, for
an arbitrary low density fluid in a stationary state, basic properties
that have been under-appreciated or overlooked, which bear upon the
collisional statistics: What is the probability distribution
$P(\cN,t)$ of the number of collisions $\cN$ suffered in equilibrium
by a tagged particle over a given duration $t$?  Conversely, what is
the probability distribution function of the free flight time,
$\PFFT(\tau)$, again for a tagged particle? Such fundamental
questions, relevant in their own right, have also consequences on the
evaluation of transport coefficients and when transposed to the
related realm of granular gases \cite{BP,BTE}, directly quantify
dissipation. Our message is that intuitive expectation fails --except
in highly untypical cases-- as far as the collisional statistics is
concerned, for an interacting fluid in and also, arguably less
surprisingly, out of equilibrium.
All results reported are new, together with the kinetic theory
techniques used.

Whereas in a dense fluid, velocity correlations and hydrodynamic
effects are responsible for a non trivial collisional statistics, one
could naively expect that in the dilute limit where collisions become
uncorrelated and molecular chaos is enforced
\cite{molchaos,Cercignani}, collisional events define a Poisson
process $\cP (\omega)$ so that $P(\cN,t)=\exp(-\omega t) (\omega
t)^{\cN}/\cN!$, where $\omega$ is the mean collision frequency
(i.e. $\langle \cN\rangle/t \to \omega$ at long times, where the
brackets denote an ensemble average). The corresponding free flight
time distribution would then be
\begin{equation}
\PFFT(\tau) \,=\, -\frac{d}{d \tau} P(\cN=0,\tau)
\,=\, \omega \,\ee^{-\omega \tau} \,. 
\label{eq:FFTnaif}
\end{equation}
However, as we shall see below --and this seems to have been ignored
in the literature \cite{rque}-- such a point of view is flawed.  In
essence, the collision frequency for a particle with velocity $\bv$
depends on $\bv$ (it generically increases with $v=|\bv|$), which in
turn induces correlations between successive collision times.  After
general considerations that encompass equilibrium and non equilibrium
steady states, we will show that the collisional statistics is
generically non Poissonian.  This is what prompted us to focus on the
simplest -analytically tractable- interactions and consider the
equilibrium hard sphere fluid for illustrative purposes.  Such a model
is one of the most useful paradigms in statistical mechanics and has
played an essential role in the development of the theory of liquids
\cite{Hansen}. It is one of the simplest system exhibiting a phase
transition.  Remarkably, it does not only provide a valuable
theoretical starting point, but also enjoys direct experimental
realizations \cite{HS}.  It is therefore surprising that such a well
studied system yields non trivial properties in a limit where little
would have been expected.  We will see that at late times, when
$\omega t\gg \cN$, the effect of such correlations is that $P(\cN,t)$
is of Poissonian form but with a renormalized frequency
$\omega/\sqrt{2}$ instead of $\omega$.  This result holds irrespective
of space dimension.
In addition, explicit and accurate results
will be reported for  the number of collision cumulants
$\langle \cN^p\rangle_c$. Our analytical predictions will be
compared to numerical simulations.

\begin{figure}
  \includegraphics[width=0.46\textwidth,clip=true]{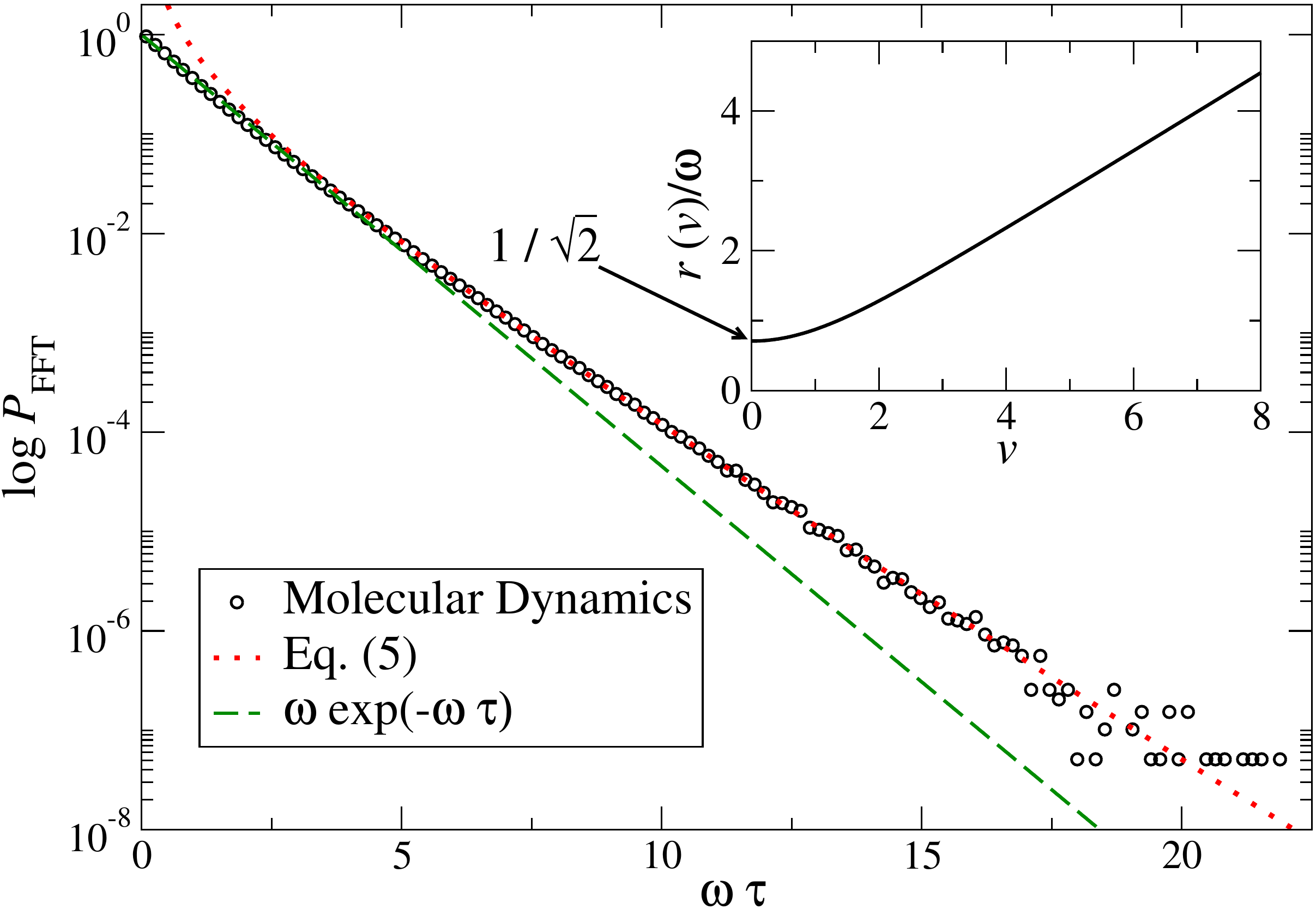}
  \caption{\label{fig:FFT} Free flight time distribution as a function
    of $\omega \tau$, on a linear-log scale, for a two dimensional
    hard disc gas ($d=2$). The circles correspond to the results of
    molecular dynamics simulations at density $\rho = 0.04
    \sigma^{-2}$ where $\sigma$ is the discs diameter, with $N=1000$ 
    particles. The dotted line
    shows the prediction of Eq. (\ref{eq:FFTroot2}) while Eq.
    (\ref{eq:FFTnaif}) is shown with a dashed line. The inset shows
    the tagged collision frequency $r(v)/\omega$ versus $v$, also for
    $d=2$. The value $1/\sqrt{2}$ at $v=0$, indicated with an arrow,
    is $d$-independent.  }
\end{figure}

For the sake of simplicity, we begin the analysis by the free flight
time distribution $\PFFT(\tau)$. The evolution of a tagged particle in
a large stationary homogeneous fluid defines, in the low density
limit, a Markov process where the transition rates can be computed
from a linearized Boltzmann equation, see e.g \cite{puglisi06}.  In
simplified situations, the velocity dependent collision rate $r(v)$
can be computed analytically; the inset of Fig. \ref{fig:FFT} shows an
illustrative example for hard spheres in equilibrium.  Velocities are
expressed here in rescaled units, and for equilibrium situations, the
velocity distribution function reads, with $d$ the space dimension
\begin{equation}
\phi(\bv) \,=\, \frac{1}{\sqrt{2\pi}^d} \, \ee^{-v^2/2}.
\label{eq:MB}
\end{equation}
Out of equilibrium, $\phi$ is a stationary measure. In any case,
the mean collision frequency $\omega$ follows from $\phi$:
\begin{equation}
\omega \,=\, \int \phi(\bv) \, r(v)\, d{\bv} = \langle r(v) \rangle\,.
\end{equation}
From the Markovian property, it follows that the {\em conditional}
probability of having a free flight time $\tau$ given a velocity $\bv$
reads $\PFFT(\tau|\bv) = r(v)\, \exp(-r(v) \tau)$.  To proceed further
and obtain $\PFFT(\tau)$ from some average of $\PFFT(\tau|\bv)$,
attention must be paid to the fact that the relevant weight to use is
not the velocity distribution $\phi(\bv)$ itself, but the velocity on
collision, $r(v) \phi(\bv)/\omega$.  The prefactor $r(v)$ biases the
distribution toward more energetic events and accounts for the fact
that in a given time interval, a particle with a larger than typical
velocity will experience more collisions. We therefore obtain
\begin{equation}
\PFFT(\tau) \, =\,
\int d\bv\,\, \frac{r^2(v)}{\omega} \,
\ee^{-r(v)\tau} \,
\label{eq:PFFT}
\phi(\bv).
\end{equation}
This expression explicitly differs from the result reported in
\cite{wiegel76}, where $r^2/\omega$ is replaced by $r$ (in other
words, the weight used in \cite{wiegel76} is $\phi$ and not $r \phi
/\omega$). To see why such an approach is incorrect, one can compute
the mean collision time $\langle \tau \rangle = \int \tau \PFFT(\tau)
\, d\tau$, that should be equal to $1/\omega$. This is indeed the case
with the distribution given in (\ref{eq:PFFT}), whereas the formula of
Ref.  \cite{wiegel76} gives $\langle \tau \rangle = \langle 1/r
\rangle$, which differs from $1/\langle r\rangle = 1/\omega$
\cite{rque10}.  More importantly, upon neglecting the $v$ dependence
of the rate $r$ (i.e. assuming $r=\omega$), the integral in
(\ref{eq:PFFT}) is readily integrated and yields expression
(\ref{eq:FFTnaif}) for $\PFFT$. It has been shown that a
$v$-independent collision rate corresponds to particles interacting
via an inverse power law pair potential with exponent $2d-2$
\cite{Cercignani}, which defines the so-called Maxwell model
\cite{Ernst}, a particularly convenient framework in kinetic
theory. Maxwell molecules are nevertheless highly untypical and for
any other fluid, $r$ depends on $v$ so that (\ref{eq:PFFT}) cannot be
exponential. We therefore conclude here that the collisional
statistics is in general non Poissonian, except for Maxwell molecules
where successive collisions turn out to be uncorrelated.  We will
clarify below the conditions for the occurrence of correlations, and
show that while the $v$ dependence of $r$ is a necessary condition for
non Poissonian behavior, it is in general not sufficient.

After the previous qualitative remarks, our goal is to quantify the
deviations for Poissonian behavior, and to this end, we hereafter
consider the prototypical hard sphere model where the frequency $r(v)$
takes the form \cite{puglisi06}:
\begin{multline}
 r(v)= \frac{\omega}{\sqrt{2}}\,
  \left( \frac{{\Mvariable{v}}^2}{d}\,
    \Mfunction{_1F_1}\left(
      \frac{1}{2},1 + \frac{d}{2},
      -\frac{{\Mvariable{v}}^2}{2}\right) \right. +\\  \left.  
     \ee^{-\frac{{\Mvariable{v}}^2}{2}} 
    \,\Mfunction{_1 F_1} \left(\frac{d-1}{2},\frac{d}{2},
      \frac{{\Mvariable{v}}^2}{2}\right) 
  \right)\,\,, 
\label{eq:rdev}
\end{multline}
where $_1F_1$ denotes a confluent hypergeometric function of the first
kind. Although a closed-form expression cannot be obtained for
$\PFFT(\tau)$ due to the lack of simplicity of the collision rate
$r(v)$, finding the large $\tau$ behavior calls for a saddle point
approximation for the integral appearing in Eq. (\ref{eq:PFFT}),
which yields
\begin{equation}
\label{eq:FFTroot2}
\PFFT(\tau) ~\stackrel{\omega \tau \gg 1}{\sim} ~\exp\left({-
\frac{\omega \tau}{\sqrt{2}}}\right) \frac{\omega}{2}
\left(1-\frac{2}{d} + \frac{\omega \tau}{\sqrt{2} d} \right)^{-d/2}.
\end{equation}
Interestingly, to leading order, $\tau$ is distributed exponentially,
as naively expected [see Eq. (\ref{eq:FFTnaif})], but with a
renormalized rate $\omega/\sqrt{2}$.  The validity of expression
(\ref{eq:FFTroot2}) is illustrated in Fig. \ref{fig:FFT}, which
displays results of numerical simulations. We have checked that the
molecular dynamics data in Fig. \ref{fig:FFT} precisely coincide with
the numerical integration of Eq. (\ref{eq:PFFT}) for all velocities
(not shown).

We also note here that very similar considerations hold for the
distribution of path lengths: for a particle with velocity $\bv$, the
free flight distance (FFD) $\ell$ travelled in a time $t$ is
$\ell=vt$, so that $P_{_{\text{FFD}}}(\ell|\bv) = v^{-1} \PFFT(\ell
v^{-1} | \bv)$.  $P_{_{\text{FFD}}}$ then follows from the counterpart
of Eq. (\ref{eq:PFFT})
\begin{equation}
\label{eq:PFFD}
P_{_{\text{FFD}}}(\ell) \,=\, \int d\bv\,\, \frac{r^2(v)}{\omega \, v}
\, \exp\left(-\frac{r(v)}{v} \ell \right) \, \phi(\bv).
\end{equation}
The resulting probability density is not a simple exponential, at
variance with a claim sometimes found in the literature (see e.g.
\cite{Hecht, BlairKudrolli}). However, a saddle point computation akin
to that leading to (\ref{eq:FFTroot2}) provides here the long distance
behavior of $P_{_{\text{FFD}}}$, which is governed by the minimum of
the function $r(v)/v$ \cite{rque2}.  This leads {\em a)} to
$P_{_{\text{FFD}}}(\ell) \propto \exp(- \omega \langle v\rangle^{-1}
\ell /\sqrt{2})$ for $\ell$ much larger than the mean free path, and
{\em b)} to the remark that for the Maxwell model with a $v$
independent rate $r$, the minimum of $r(v)/v$ is reached for
$v\to\infty$ and vanishes, which leads to a non-exponential large
$\ell$ limiting behavior \cite{prep}.

We now turn to the related but more complex question of the number of
collisions. We introduce the joint probability $f(\bv, \cN, t)$ of
having velocity $\bv$ and having suffered $\cN$ collisions in a time
window $t$, for our tagged particle.  The corresponding time
evolution, again in the dilute limit, follows from the linear
Boltzmann-like equation
\begin{multline}
  \p_t f(\bv_1,\cN,t)=\int \dd \bv_2 \int \dd \bs (\bv_{12} \cdot
  \bs) \theta (\bv_{12} \cdot\bs)\\
  \left[f(\bv_1^{**},\cN-1,t) \phi(\bv_2^{**}) - f(\bv_1,\cN,t)
    \phi(\bv_2) \right] \,\,,
  \label{eq:boltz}  
\end{multline}
where $\theta$ is the Heaviside function, $\bv_{12}$ is the relative
velocity, $\bs$ is a unit vector and the $**$ superscript refers to
pre-collisional velocities: $\bv_1^{**}=\bv_1-(\bv_{12}\cdot\bs)\bs$
and $\bv_2^{**}=\bv_2+(\bv_{12}\cdot\bs)\bs$. The latter equation
encodes a full description of the collisional statistics for the
tagged particle, in the low density limit. Here again we stress that
such an analytical approach can be extended to other interaction
potentials, leading to the already mentioned remark that, apart from
Maxwell molecules, it does not admit a Poisson solution.  Moreover,
the above equation does not admit a stationary solution in the long
time limit, due to the time dependent behavior of the collision number
$\cN$. More precisely, we expect the large time dependence to be
exponential, as a consequence of the linear character of the
equation. For analytical progress, it turns convenient to introduce
the generating function $\hf$ through
\begin{equation}\label{dwell}
  \hf(\bv,\lambda,t) = \sum_{\cN = 0}^{\infty} \ee^{-\lambda \cN}
  f(\bv, \cN,t) \,.
\end{equation}
Of course, upon summing (\ref{eq:boltz}) over all possible values of
$\cN$, in the equilibrium state one recovers $\sum_{\cN=0}^{\infty}
f(\bv,\cN) = \hf(\bv,0,t)=\phi(\bv)$, the Maxwell-Boltzmann
distribution (\ref{eq:MB}). It then appears that the cumulant
generating function $\mu(\lambda)$, such that in the large time limit
\begin{equation}
  \langle \cN^p \rangle_c \stackrel{\omega t \gg 1}{\sim} 
  t (-1)^p \left. \frac{\p^p \mu}{\p \lambda^p} 
  \right|_{\lambda=0}\,\,,
\label{eq:cumul}
\end{equation}
is the largest eigenvalue of an evolution operator that
straightforwardly follows from (\ref{eq:boltz}) \cite{prep}.
Furthermore, $\mu(\lambda)$ is directly related to the large time
behavior of $P(\cN,t)$ through its large deviation function $\pi$,
defined as
\begin{equation}
P(\cN,t) ~\stackrel{\omega t \gg 1}{\sim} ~ \ee^{t \pi(n)}
\end{equation}
where $n=\cN/t$. Indeed, $\pi$ is the Legendre transform of $\mu$:
\begin{equation}
\pi(n)=\min_{\lambda} (\mu(\lambda) + \lambda n)\,\,.
\label{eq:pimin}
\end{equation}
The quantity $\mu(\lambda)$ therefore bears an important physical
information, and has been the technical focus of our study.

From perturbation theory, we have obtained the behavior of $\mu$ at
large $\lambda$ in the form
\begin{equation}
\label{mu_largelambda}
\mu(\lambda) \sim  \frac{\omega}{\sqrt{2}} (\ee^{-\lambda} -1) 
+ {\cal O} (\ee^{-2 \lambda})\,\,,
\end{equation}
which implies that
\begin{equation}
  P(\cN,t) \sim \frac{\ee^{-\frac{\omega t}{\sqrt{2}}}}{\cN !}
  \left(\frac{\omega t}{\sqrt{2}}\right)^{\cN}\,\,,\qquad 
  \textrm{for $\cN \ll \omega t$}\,\,.
\label{eq:blabla}
\end{equation}
For this Poissonian behavior $\cP(\omega/\sqrt{2})$, the large
deviation function easily follows:
\begin{equation}
\pi(n) =  n - n \log (n \sqrt 2/\omega) -\omega/\sqrt{2} \ .
\label{eq:pinpoisson}
\end{equation}
We note that Eq. (\ref{eq:blabla}) with $\cN=0$ is compatible with the
time integral of the leading exponential order of free flight time
distribution given in (\ref{eq:FFTroot2}), as it should.

\begin{figure}
  \includegraphics[width=0.46\textwidth,clip=true]{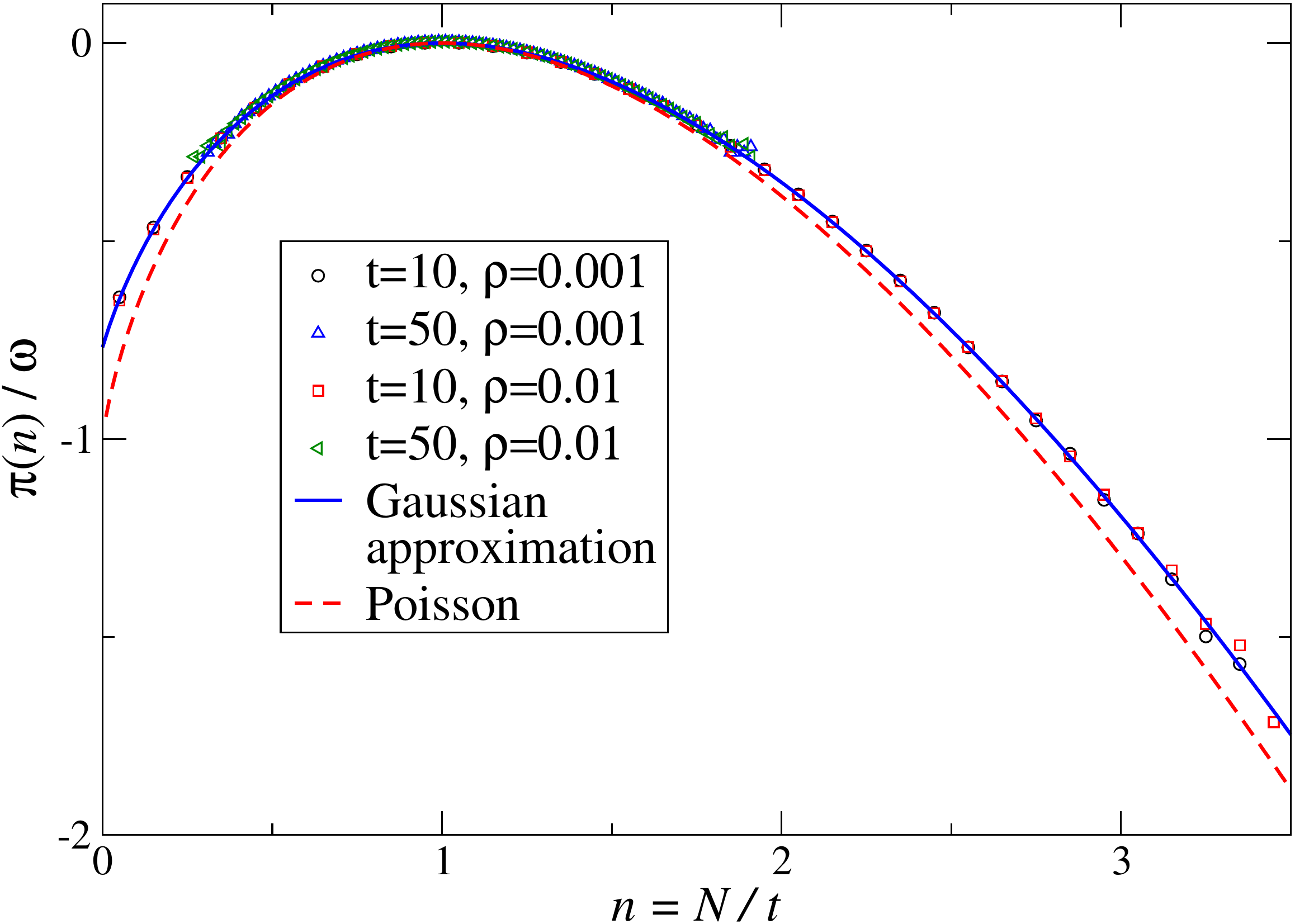}
  \caption{ Large deviation function $\pi(n)$ characteristic of the
  long time behavior of the probability $P(\cN,t)$ of suffering $\cN$
  collisions in a time $t$ (and defined by $\pi(\cN/t) \sim t^{-1}
  \log P(\cN,t)$ for $\omega t \gg 1$). The symbols correspond to
  molecular dynamics measures for a hard disc system with $N=1000$
  particles at two different but low densities.  The solid line shows
  the Gaussian result following from Eqs. (\ref{mu_noa2}) while the
  dashed line is the large deviation function $\pi(n) =
  n-n\log(n/\omega) -\omega$ associated to the Poisson law
  $\cP(\omega)$. On the graph, the time scale has been set by the
  choice $\omega=1$.
  \label{fig:ldf} }
\end{figure}

However, the dependence embodied in Eqs. (\ref{mu_largelambda}) and
(\ref{eq:blabla}) follows from a large $\lambda$ expansion and only
holds for $\cN \ll \omega t$ (hence $n/\omega \to 0$). It is therefore
not indicative of the typical behavior, for which it is more useful to
consider the low $\lambda$ limit. We then dwell on the remark made
after (\ref{dwell}) that for $\lambda=0$, we have
$\hf(\bv,0,t)=\phi(\bv)$, which leads to the approximation scheme
where $\hf(\bv,\lambda,t)$ is taken as a Gaussian with temperature
$T(\lambda)$ and the requirement that $T(0)=1$ \cite{rque3}. This
approximation is the lowest order of a more systematic expansion (see
\cite{prep}), but it provides a useful and reasonably accurate
information given its simplicity. The idea is to project the Boltzmann
equation (\ref{eq:boltz}) governing the evolution of
$\hf(\bv,\lambda,t)$ onto the first two velocity moments. This leads
to
\begin{subequations}
\label{mu_noa2}
\begin{equation}
  \mu(\lambda)=- \frac{\omega}{\sqrt{2}} (1-\ee^{-\lambda})
  \sqrt{1+\frac{T(\lambda)}{T_0}}\,\,,
\end{equation}
\begin{equation}
  \hbox{with } ~ T(\lambda)=\frac{\sqrt{2}
  T_0}{\sqrt{1+\ee^{\lambda}}}\,\,.~~~~~~~
\end{equation}
\end{subequations} 
It can be checked that $\mu(\infty)=-\omega/\sqrt{2}$, as implied by
Eq. (\ref{mu_largelambda}). The corresponding first three cumulants
follow from (\ref{eq:cumul}) and read, irrespective of dimension $d$
\begin{subequations}
  \begin{align}
    \frac{\langle \cN \rangle_c}{\omega t} & = 1 \,\,  \\
    \frac{\langle \cN^2 \rangle_c}{\omega t} & = \frac{9}{8}\,\,\\
    \frac{\langle \cN^3 \rangle_c}{\omega t} & = \frac{289}{256}\,\, .
\end{align}
\label{eq:Gauss}
\end{subequations}
These values are compared to molecular dynamics simulation data in
Table \ref{table:1}. Moreover, the large deviation function $\pi(n)$
of collisions, which follows from (\ref{eq:pimin}), appears to be in
excellent agreement with its molecular dynamics counterpart, see Fig.
\ref{fig:ldf}

\begin{table}
\caption{\label{tab:cumulants} Cumulants for the number of collisions
    $\cN$ from molecular dynamics simulations (performed on a two
    dimensional system with reduced density $\rho \sigma^2 = 0.04$),
    and comparison with both the Poisson $\cP(\omega)$ result and
    Gaussian approximation (\ref{eq:Gauss}). }
  \begin{ruledtabular}
\begin{tabular}{lcccr}
  &$\langle \cN \rangle_c /t$ & $\langle \cN^2 \rangle_c /t$ & 
  $\langle \cN^3 \rangle_c /t$\\
  \hline
  $\omega t=10$  & 1. & 1.123 & 1.129 \\
  $\omega t=50$  & 1. & 1.135 & 1.105   \\
  Poisson & 1  & 1       & 1       \\
  Gaussian& 1  & 1.125   & 1.129 \\
\end{tabular}
\end{ruledtabular}
\label{table:1}
\end{table}

At this point, it proves instructive to consider the Lorentz gas with
only one mobile particle and a collection of spherical fixed
scatterers.  One readily gets $r(v) \propto v$, but the collisional
statistics is nevertheless Poissonian. The reason is that the velocity
modulus of the mobile particle is constant along the trajectory: in
other words, there is no thermalization.  It therefore appears that
non Poissonian behavior arises from two key properties, that induce
collisional correlations: first, the collision rate depends on the
velocities, and, second, the particle thermalizes to some stationary
non singular measure. These criteria also apply out of equilibrium.

In conclusion, we have shown that the statistics of the number of
collisions $\cN$ experienced by a tagged particle in a low density
homogeneous and stationary fluid, in or out of equilibrium, is a
subtler quantity than it might seem. Our general statements have been
illustrated with the hard sphere fluid.  For the distributions of both
$\cN$ and the related free flight time --for which several incorrect
results may be found in the literature--, we have quantified the
corresponding non Poissonian behavior which follows from the simple
physical ingredient that a particle with a high velocity statistically
collides more often than a typical particle. A key quantity in the
theoretical analysis is the cumulant generating function
$\mu(\lambda)$, which can be computed explicitly for large $\lambda$
and approximately for small $\lambda$.  The Gaussian ansatz worked out
here can be considered as the lowest order of a systematic expansion.  
The resulting analytical predictions have been
confronted against molecular dynamics numerical simulations, with a
very good agreement.  These numerical results show that the deviations
from Poisson behavior $\cP(\omega)$, although not dramatic --which may
be the reason why they are under-documented in the literature-- are
nevertheless clearly observable. In particular, we have obtained the
{\it a priori} surprising result that for long times, the distribution
of $\cN$ is Poissonian, but with a ``dressed'' rate $\omega/\sqrt{2}$.
Conversely, the distribution of free flight time $\tau$ is
exponential, with a behavior $\propto \exp(-\omega \tau /\sqrt{2})$.

{\em Acknowledgements} We would like to thank J. Piasecki, J.M.J. van
Leeuwen, M.H. Ernst, D. Frenkel and H. van Beijeren for useful
discussions.  This work was supported by the French Ministry of
Education through a JCJC ANR grant.



\begin{thebibliography}{10}
\bibitem{rque2} Conversely, the minimum of $r(v)$ is the relevant
quantity for the long time behavior of Eq. (\ref{eq:PFFT}); it is
reached at $v=0$ (see the inset of Fig. \ref{fig:FFT}), where
$r(0)=\omega/\sqrt{2}$.  This explains the $\sqrt{2}$ factors
appearing in Eq. (\ref{eq:FFTroot2}).
\bibitem{rque3} More precisely, it can be shown that at late times,
  $\hf(\bv,\lambda,t)$ takes the asymptotic form $\exp[\mu(\lambda) t]
  \tilde f(\bv,\lambda)$, so that the time dependence factorizes from
  the $\bv$ variable, which explains why $T(\lambda)$ is
  $t$-independent. The requirement that $\mu(0)=0$ ensures that
  $\hf(\bv,0,t)$ is also time independent.
\end{thebibliography}
\end{document}